\documentclass[aps,prl,twocolumn,preprintnumbers]{revtex4}

\usepackage{amsmath,graphicx,color,amssymb}

\def\sgra{Sgr A${}^{*}$}

\def\beq{\begin{equation}}
\def\eeq{\end{equation}}

\def\avg#1{\langle #1 \rangle}

\newcommand{\nc}{\newcommand}
\nc{\Tr}{\mbox{Tr}}
\nc{\hc}{\mbox{H.c.}}
\nc{\ev}{\;\mathrm{eV}}
\nc{\mev}{\;\mathrm{MeV}}
\nc{\gev}{\;\mathrm{GeV}}
\nc{\infinity}{\infty}

\def\apjs{Astrophys. J. Suppl.}

\begin{document}

\title{Galactic Center Gamma-Ray Excess from Dark Matter Annihilation:\\
Is There A Black Hole Spike?}

\author{Brian D. Fields}
\author{Stuart L. Shapiro}
\author{Jessie Shelton}
\affiliation{Departments of Physics and of Astronomy, University of Illinois at
  Urbana-Champaign, Urbana, IL 61801, USA}

\begin{abstract}
  If the supermassive black hole \sgra\ at the center of the Milky Way grew
  adiabatically from an initial seed embedded in an NFW dark matter
  (DM) halo, then the DM profile near the hole has steepened into a
  spike.  We calculate the dramatic enhancement to the gamma ray flux
  from the Galactic center (GC) from such a spike if the 1-3 GeV
  excess observed in {\em Fermi} data is due to DM annihilations.  We
  find that for the parameter values favored in recent fits, the point
  source-like flux from the spike is 35 times greater than the flux
  from the inner $1^\circ$ of the halo, far exceeding all {\em Fermi}
  point source detections near the GC.  We consider the
  dependence of the spike signal on astrophysical and particle
  parameters and conclude that if the GC excess is due to
  DM, then a canonical adiabatic spike is disfavored by the data.  We discuss
  alternative Galactic histories that predict different spike signals,
  including: (i) the nonadiabatic growth of the black hole, possibly
  associated with halo and/or black hole mergers, (ii) gravitational
  interaction of DM with baryons in the dense core, such as heating by
  stars, or (iii) DM self-interactions.  We emphasize that
  the spike signal is sensitive to a different combination of particle
  parameters than the halo signal, and that the inclusion of a spike
  component to any DM signal in future analyses would provide novel
  information about both the history of the GC and the
  particle physics of DM annihilations.

\end{abstract}

\maketitle


The indirect detection of high-energy particles
originating in dark matter (DM) 
annihilations or decays is a cornerstone in
the search for DM (see, e.g.,~\cite{Buckley:2013bha} for a
recent review).  Annihilations in the Galactic halo may lead to a
signal from the Galactic center (GC) at rates that are
observable by the current generation of high-energy experiments.  The
excess of $\sim 1-3$ GeV gamma rays from the inner few degrees of the
GC observed in {\it Fermi} telescope data may be such a
signal
\cite{Goodenough:2009gk,Hooper:2010mq,Hooper:2011ti,Abazajian:2012pn,Hooper:2012sr,Hooper:2013rwa,Gordon:2013vta,Huang:2013pda,Abazajian:2014fta,Daylan:2014rsa}.
A recent analysis~\cite{Daylan:2014rsa} of the {\it Fermi} data with
improved angular resolution~\cite{Portillo:2014ena} has sharpened the case
for a DM interpretation of the excess, demonstrating a clear preference
for a component of emission from a spherically symmetric, extended
source with an energy spectrum apparently independent of position.

A supermassive black hole (SMBH) exists at the site of
\sgra~\cite{GenEG10,Ghez.etal08}.  Such an object should steepen the
DM density profile in its sphere of influence. If the SMBH grows
adiabatically from a smaller seed, the resulting density spike yields
a strong enhancement of any DM annihilation
signal~\cite{Gondolo:1999ef,Mer04,GneP04}.  Here we construct a
canonical GC model containing an adiabatic density spike. 
We adopt the
best-fit halo and particle parameters found in \cite{Daylan:2014rsa}
to calculate the expected gamma-ray flux and spectrum from both the
spike and ambient Navarro-Frenk-White (NFW~\cite{NavFW96}) halo regions.  
We find that the expected flux
from the spike in our canonical model considerably exceeds the flux
from any of the pointlike sources near the GC cataloged by {\em Fermi}. To reconcile the discrepancy we discuss plausible
alternatives for our model for the Galactic halo and spike, as well as
different allowed choices for the DM particle physics properties. We
emphasize the importance of incorporating both SMBH spike and halo
components in future analyses of GC annihilation fluxes.

The effects of adiabatic black hole-induced DM spikes on the
isotropic diffuse gamma-ray background have been considered in
\cite{Belikov:2013nca}, while the effects on gamma rays of potential spikes in dwarf galaxies were
studied in \cite{Gonzalez-Morales:2014eaa}.

\vspace{0.2cm}
\noindent {\it{\bf The Canonical Model: Galactic Center Halo and Spike.}}
We assume that the DM is collisionless and adopt a cuspy, spherical,
DM matter density profile obeying a NFW
like-halo profile.  A DM density spike due to the
presence of the central, supermassive black hole (SMBH) \sgra\ 
forms inside the radius of gravitational influence of the SMBH, $r_h =
M/v_0^2$ ($G\equiv 1$). Here $M$ is the mass of the hole and $v_0$ is the (1-d)
velocity dispersion of DM in the halo outside the spike. We assume
that the spike forms in response to the adiabatic growth of the
SMBH~\cite{Pee72a,Gondolo:1999ef,Mer04,GneP04}.  Our adopted DM
density profile may then be approximated by connected power-law
profiles of the form
\begin{eqnarray}
\rho(r) &=& 0, \ \ \ r < 4M \ \ ({\rm capture \ region}), \\
&=& \frac{\rho_{\rm sp}(r)\rho_{\rm in}(t,r)}
{\rho_{\rm sp}(r) + \rho_{\rm in}(t,r)},  
\ 4M \leq r < r_b \ \ ({\rm spike}), \nonumber \\
&=& \rho_b(r_b/r)^{\gamma_{c}}, \ \ \ \ \ \ r_b \leq r < R_H \ \ \
({\rm cusp}), \nonumber \\
&=& \rho_H (R_H/r)^{\gamma_H}, \ \ \ R_H \leq r \ \ \ 
({\rm outer \ halo}), \nonumber
\end{eqnarray}
where $r_b = 0.2 r_h$, 
$\rho_b = \rho_D (D/r_b)^{\gamma_{c}}$, 
$\rho_{\rm sp}(r)=\rho_b(r_b/r)^{\gamma_{sp}}$,
$\rho_{\rm in}(t,r)=\rho_{\rm ann}(t)(r/r_{\rm in})^{-\gamma_{\rm in}}$,
and 
$\rho_H = \rho_D (D/R_H)^{\gamma_{c}}$. 
Here $\rho_D$ is the DM density in the solar neighborhood, a distance
$D$ from the GC. The density $\rho_{\rm ann}$ is the
so-called DM ``annihilation plateau'' density $\rho_{\rm ann} =
m_{\chi}/{\avg{\sigma v} t}$, reached by $\rho_{\rm sp}(r)$ in the
innermost region of the spike at $r=r_{\rm in}$, where $m_{\chi}$ is
the mass of the DM particle, $\sigma$ is the annihilation cross
section, $v$ is the relative velocity and $t=t_{\rm ann}$ is the
lifetime over which annihilations have occurred ($\approx$ the age of
the SMBH).  The radius $R_H$ denotes the outer halo, and joining it
onto the halo cusp yields $\rho_H = \rho_D (D/R_H)^{\gamma_c}$.

For the velocity dispersion profile, assumed isotropic, we
take 
\begin{eqnarray}
v^2(r) &=&  \frac{M}{r} \frac{1}{1+\gamma_{in}} 
\left[
1 + \frac{r}{r_{\rm in}}
\left( \frac{\gamma_{in}-\gamma_{sp}}{1+\gamma_{sp}} \right)
\right], \nonumber \\
&& \ \ \ \ \ \ \ \ \ \ \ \ \ \ \ \ 4M \leq r < r_{\rm in} \ \
({\rm inner \  spike}), \label{eq:velocity}\\
&=& \frac{M}{r} \frac{1}{1+\gamma_{sp}},  \ \ \ 
r_{\rm in} \leq r < \frac{r_h}{1+\gamma_{sp}} \ \ ({\rm outer \  spike}), 
 \nonumber \\
&=& v_0^2 = {\rm const},   \ \ \frac{r_h}{1+\gamma_{sp}}\leq r 
\ \ ({\rm cusp \ \& \ outer \ halo}). \nonumber
\end{eqnarray} 
Here we take the dispersion in the DM halo to be nearly constant
outside the spike and match it onto an approximate, piece-wise
continuous solution of the Jeans equation in the spike.  We neglect
relativistic corrections near the SMBH (see~\cite{SadFW13}) but set
the DM density to zero inside $4M$, the radius of marginally bound
circular orbits and the minimum periastron of all parabolic orbits
about a Schwarzschild black hole.

We adopt the following parameter values for our canonical Milky Way DM
halo and adiabatic spike:
$M = 4 \times 10^6 {\rm M_{\odot}}$~\cite{GenEG10,Ghez.etal08},
$\rho_D = 0.008 \pm 0.003~ {\rm M_{\odot}~pc^{-3}} = 0.3 \pm 0.1{\rm
  ~GeV ~cm^{-3}}$~\cite{BovT12}, $v_0 = 105 \pm 20 {\rm
  ~km~s^{-1}}$~\cite{Gultekin:2009qn}, $D = 8.5$ kpc~\cite{Daylan:2014rsa},
$R_H = 16$ kpc~\cite{NesS13}, and $t_{\rm ann} = 10^{10}$ yrs. With
these parameters we find a spike radius of $r_h = 1.7$ pc,
corresponding to $0.012^{\circ}$, well below the resolution of {\it
  Fermi}~\cite{fermiangular} and even below the resolution envisioned for
successor telescopes such as {\it Gamma-Light}~\cite{gammalight}.  The
inner boundary of the spike is at $r = 4M = 6 \times 10^6$ km.

We note that $\gamma_c = 1$
and $\gamma_H = 3$ are the standard NFW values for the inner and outer
halo regions, respectively.  For our canonical choice we instead take
$\gamma_c = 1.26$, the best-fit value reported in
Ref.~\cite{Daylan:2014rsa} and consider variations about this
value. We adopt $\gamma_H = 3$. For a spike of collisionless matter
that forms about an adiabatically growing SMBH we have $\gamma_{sp} =
(9-2\gamma_c)/(4-\gamma_c)$, which yields $\gamma_{sp} = 2.36$ for our
canonical choice. We note that for $0 \leq \gamma_c \leq 2$ the spike
power-law $\gamma_{sp}$ varies at most between 2.25 and
2.50~\cite{Gondolo:1999ef}. In the innermost region, annihilations
weaken (but do not flatten) an isotropic spike, whereby $\gamma_{\rm
  in} = 1/2$~\cite{Vas07}.

We note that the effects of baryons may lead to significant departures
in the above profiles, but this issue has not been settled.  For
example, the heating of DM by gravitational scattering off stars could
lower the DM density inside the spike considerably over 10
Gyr~\cite{Mer04,GneP04}.  The original calculations suggested that
this heating would drive $\gamma_{sp}$ down to $ \sim 1.8$ in 10 Gyr,
ultimately reaching a final equilibrium profile in $\gtrsim 20$ Gyr
with $\gamma_{sp}=3/2$.  Such scattering could even replenish the
annihilated DM in the spike and lift the annihilation plateau,
$\rho_{\rm ann}$.  However, these early results were based on the
belief that there exists a steeply rising density of stars in the
Milky Way inside $r_h$. Careful observations in 2009 showed that the
dominant old, late-type stars in fact had a distribution that was
flat, or even decreasing, toward the GC (see \cite{Mer13}
for a discussion), resulting in long relaxation timescales and
corresponding heating timescales well above 10 Gyr. But a shallow
spike profile with $\gamma_{sp} = 3/2$ could also arise inside cored
halos, which might form if, for example, the halo underwent mergers,
cannibalism, or other cataclysmic dynamical changes such as strong
supernova feedback that led to large potential field fluctuations
prior to the adiabatic growth of \sgra.  There are other possibilities that could weaken the spike:
the formation history of the Milky Way and \sgra, the influence
of an unseen distribution of stars (e.g.~compact objects) in the spike,
or DM self-interactions~\cite{ShaP14}. In the opposite limit
from adiabatic growth, were the SMBH to appear instantaneously (e.g.,
via a rapid sub-halo merger) the spike would be quite shallow:
$\gamma_{sp} = 4/3$~\cite{Ull01B}.  The sudden formation of a seed SMBH
followed by its adiabatic growth would lead to intermediate
slopes. Off-center ($\gtrsim 50$~pc) formation of the SMBH
might flatten the spike entirely unless appreciable
growth occurred after the hole's arrival at the GC.

For our canonical DM annihilation cross section and mass we adopt as a
default model the reference point of Ref.~\cite{Daylan:2014rsa}: a
self-conjugate DM particle with mass $m_{\chi} = 35.25$ GeV
annihilating to $b\bar b$ with a cross-section $\avg{\sigma v} = 1.7
\times 10^{-26} {\rm cm}^3{s}^{-1}$, strikingly close to the
expectation for a thermal relic origin of DM.  For this particle
model, the annihilation plateau is $\rho_{\rm ann}=1.7\times
10^{8}{\rm ~M_{\odot}~pc^{-3}} = 6.6\times 10^{9}{\rm ~GeV ~cm^{-3}}
$, reached at a radius $r_{\rm in} = 3.1\times 10^{-3}$ pc in the
spike.  The GC excess can be fitted by DM models with a range of
possible masses and cross sections
\cite{Hooper:2012cw,Daylan:2014rsa,Lacroix:2014eea,Boehm:2014bia,Ko:2014gha,Abdullah:2014lla,Martin:2014sxa,Berlin:2014pya},
further considered below.

We take the DM annihilation cross section to be velocity-independent,
$\avg{\sigma v} = const$, i.e., dominated by $s$-wave interactions;
this choice renders the radial dependence of the velocity dispersion
of Eq.~\ref{eq:velocity} irrelevant for the DM annihilation signal.
At the inner boundary of the spike velocity dispersions reach $v\sim
c/5$, nearing the velocity range that pertains during thermal
freezeout of a cold relic.  Terms beyond the leading $s$-wave
contribution to the annihilation cross section could thus be
important. However, the dominant contribution to the spike signal
comes from the region $r\approx r_{\rm in}$, where the velocity
dispersion is still small: in our canonical model, $v(r_{\rm
  in})\approx 1300 \,\mathrm{km/sec}\ll c$. Thus as long as $s$-wave
contributions are present, the region around $r_{\rm in}$ will still
provide most of the spike signal and contributions from higher-order
terms in the velocity expansion will remain small.

\vspace{0.2cm}

\noindent {\it{\bf Gamma-Ray Flux from the Canonical Adiabatic Spike.}}
The differential gamma-ray (number) emissivity from (self-conjugate)
DM annihilations is 
\beq
q_E(r) 
 = \frac{\rho(r)^2}{4\pi m_\chi^2}  \frac{\avg{\sigma v}}{2} \frac{dN_\gamma}{dE_\gamma}.
\eeq
The differential number flux on earth from annihilations 
within an angle $\theta$ is then
\beq
\frac{d\Phi}{dE} (<\theta) = \int  I_E  \ \cos \theta \ d\Omega
= \int_0^\theta  \cos \theta \ d\Omega \int_{\rm los} q_E \left(r \right)\ ds \ \ .
\eeq
where the inner integral employs the intensity along the line-of-sight at
an angle $\theta$ from the GC,
and where $r(\theta,s)= \sqrt{D^2 + s^2 - 2 s D \cos \theta}$.
%

\begin{figure}
\includegraphics[width=\linewidth]{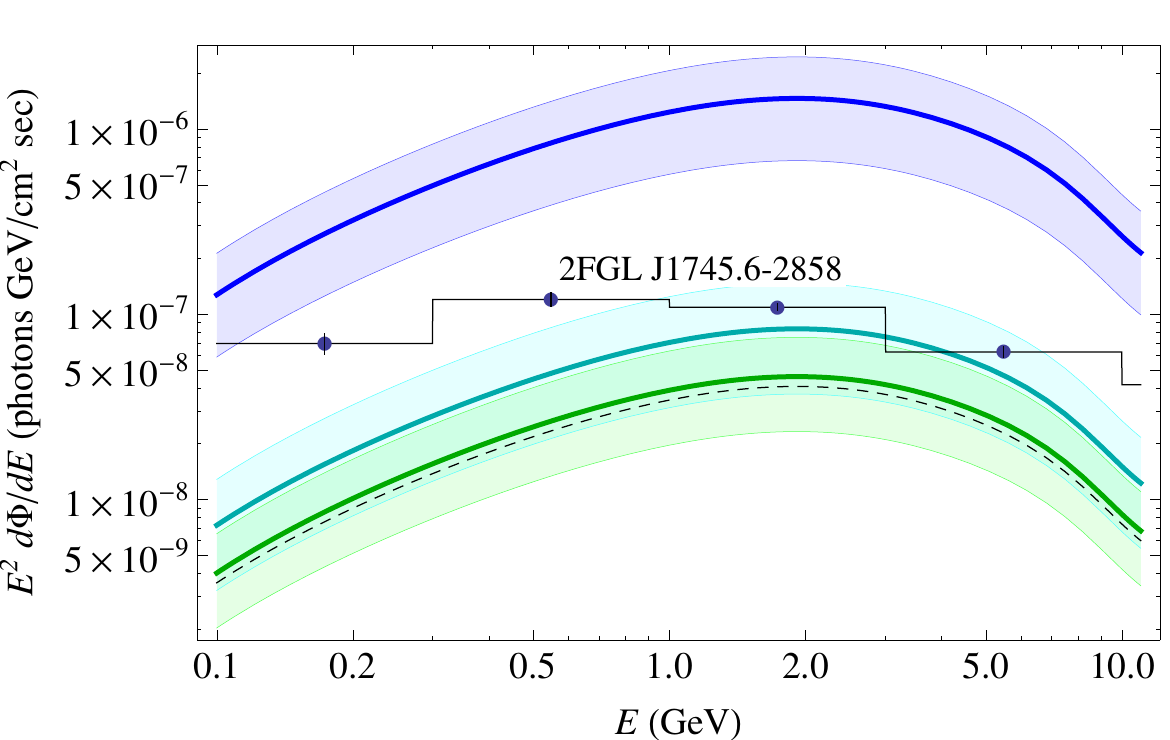} 
\caption{ GC gamma-ray flux.  Points are the {\em
    Fermi} source 2FGL J1745.6-2858 near \sgra~
  \cite{Fermi-LAT:2011iqa}.  Curves show the DM annihilation
  signal from the spike and halo
  \cite{Daylan:2014rsa} within $1^\circ$.  Top: 
  the canonical adiabatic spike plus halo
  ($\gamma_c = 1.26$, $\gamma_{sp} = 2.36$).  Middle: same as
  top, but for $\gamma_{sp} = 1.8$.  Bottom: same as top,
  but for $\gamma_{sp} = 1.5$. The dashed black line shows 
  the inner $1^\circ$ of the halo alone.  Shaded
  bands vary the halo index
  between $1.2 \leq \gamma_c \leq 1.3$.  }
\label{fig:spectrum}
\end{figure}

The spike signal 
appears as a point source at the GC, superimposed upon
the smooth halo annihilation signal as well as any astrophysical
emission from the direction of the center.  Thus the spike and halo
signals will both contribute within one point spread function (PSF) at
the center.  To approximate this smoothing we will compare the spike
flux to the integrated halo flux within $1^\circ$ \footnote{
  Front-converting photons in the {\em Fermi} LAT have 95\%
  containment angles that grow from $\sim 0.2^\circ$ at 30 GeV to
  almost $2^\circ$ at 1 GeV \cite{fermiangular}. }.

The 2nd {\em Fermi} point source catalog \cite{Fermi-LAT:2011iqa}
lists four sources within $1^\circ$ of the GC, the
brightest of which has a location consistent with that of \sgra.
This source, 2FGL J1745.6-2858, has a flux $\Phi(1-100 \ {\rm GeV})
= (7.74 \pm 0.20) \times 10^{-8} \ \mathrm{/cm^{-2} \ s^{-1}}$,
and is within 1.79 arcmin of \sgra.  We take this source as the
candidate for the signal containing the spike emission.

The {\em Fermi} measurements of the point source flux likely include
emission from the many astrophysical sources towards and near the
GC in addition to any DM signal.  Moreover, the
DM spike signal will in general contain a contribution from
the halo emission that was not included in the {\em Fermi} point
source analysis; groups that include a component 
$\propto \rho_{\rm NFW}(r)^2$ in their models of the GC
find that the flux associated to the \sgra point source is
substantially reduced, though the degree of reduction depends on the
details of the
fit~\cite{Abazajian:2012pn,Gordon:2013vta,Abazajian:2014fta}.  We thus
regard the {\em Fermi} spectrum of 2FGL J1745.6-2858 as an {\em upper
  limit} to the annihilation signal from the GC.

Fig.~\ref{fig:spectrum} shows our fiducial predictions for the
GC DM annihilation gamma-ray spectrum (curves), as well
as {\em Fermi} observations (points).  We include the flux from the
innermost $1^\circ$ of the smooth DM halo along with the contributions
from three possible spikes: the canonical adiabatic spike, the
limiting equilibrium spike from stellar heating with $\gamma_{sp} =
1.5$, and an intermediate spike with $\gamma_{sp} = 1.8$.  
We conservatively include primary photons only; 
secondary photons arising from the interaction of
DM annihilation products with gas, dust, and magnetic fields in the
GC would further increase the flux at low energies.

With the best-fit particle and halo parameters from
Ref.~\cite{Daylan:2014rsa}, the spike
emission lies more than an order of magnitude above the observed point
source emission.  The
energy-integrated flux due to the spike is $\Phi_{\rm spike}
(1-100\gev)= 1.1\times 10^{-6} \ \rm cm^{-2} \ s^{-1}$, while the
inner $1^{\circ}$ contributes only $\Phi_{\rm
  halo}(1-100\gev)=3.2\times 10^{-8} \ \rm cm^{-2} \ s^{-1}$.  The
canonical halo signal inside $1^\circ$ lies {\em below} the {\em
  Fermi} data for $\gamma_c = 1.26$, so the mismatch is entirely due
to the black hole spike signal. 

Thus, {\em emission from a canonical adiabatic black hole spike is
  inconsistent with a DM interpretation of the GC gamma-ray
  excess given the results of Ref.~\cite{Daylan:2014rsa}}.

\begin{figure}
\includegraphics[width=\linewidth]{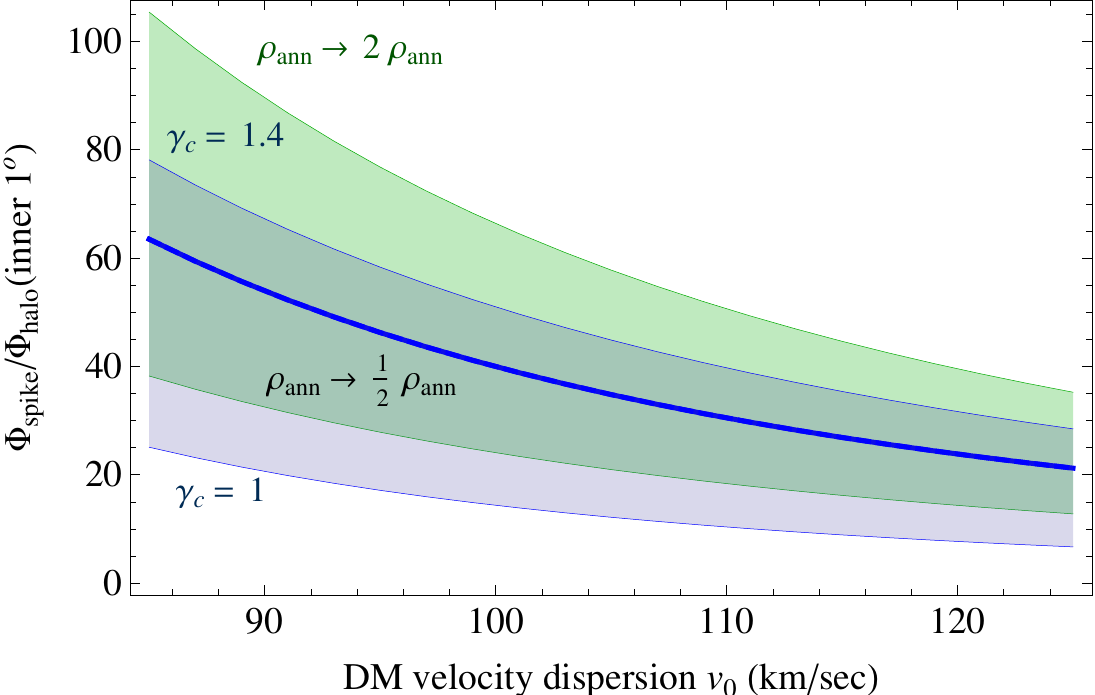}
\caption{Ratio of the canonical adiabatic spike flux to
  the flux of the inner $1^\circ$ of the halo.  The solid blue line indicates
  the predictions from the best-fit parameters of
  Ref.~\cite{Daylan:2014rsa}.  The blue
  band shows the effect of varying $1 \le \gamma_c \le 1.4$.  The
  overlapping green band shows the effect of varying the annihilation
  plateau $\rho_{\rm ann}$.
  }
\label{fig:spikehaloratios}
\end{figure}

The steep power law dependence of the canonical adiabatic black hole
spike makes it a very bright signal, allowing us to reach this sharp
conclusion. However it also renders the signal sensitive to variations
in the Galactic halo parameters.  For example, the spike signal varies
significantly with the DM velocity dispersion $v_0$, as we demonstrate
in Fig.~\ref{fig:spikehaloratios}. The velocity dispersion determines
the size of the spike through $r_h$ which, in turn, establishes the
radius $r_{\rm in}$ at which $\rho_{sp}=\rho_{\rm ann}$ and near which
most of the emission emanates.

The spike signal grows more rapidly than the halo signal as the halo
index $\gamma_c$ increases, owing to the increase in $r_{\mathrm{in}}$
when the spike grows on top of a larger initial density at the GC.
Fig.~\ref{fig:spikehaloratios} shows the result of varying $1\leq
\gamma_c\leq 1.4$, 
as well as the effect of varying $\rho_{\rm ann} = m/\avg{\sigma v}
t$.  In all cases the spike is more than 5 times brighter than the
inner $1^\circ$ of the halo.

\begin{figure}
\includegraphics[width=\linewidth]{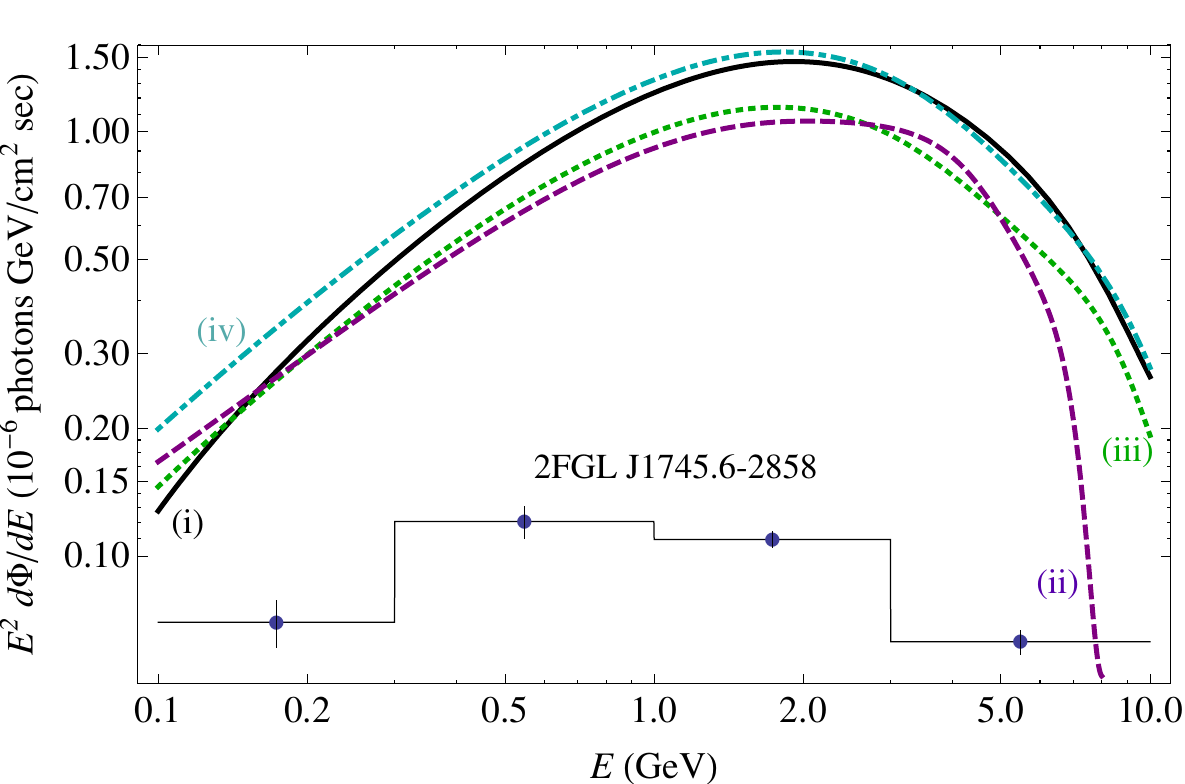}
\caption{Canonical adiabatic spike signals for different
  particle models of the GC excess.  Shown are the
 models (i)--(iv) discussed in the text.
 }
\label{fig:diffpart}
\end{figure}

Because of its dependence on $\rho_{\rm ann}$, the spike signal probes
a {\em different} combination of particle physics parameters than does
the halo signal alone.  Thus particle models that yield the same
prediction for the halo flux will predict different values of the
spike-to-halo ratio, as illustrated in Fig.~\ref{fig:diffpart}. Here
we show predicted spike signals for some representative particle
models that have been advanced for the excess: (i) our reference model
\cite{Daylan:2014rsa}; (ii) a 9 GeV self-conjugate particle
annihilating to $\tau^+\tau^-$ 80\% of the time and $b\bar b$ 20\% of
the time with cross-section $\avg{\sigma v} = 0.7\times
10^{-26}\mathrm{cm^3/s}$ \cite{Daylan:2014rsa}; (iii) a 22 GeV
non-self-conjugate particle annihilating to 16 GeV kinetically mixed
dark vector bosons with cross-section $\avg{\sigma v} = 3.4\times
10^{-26}\mathrm{cm^3/s}$ \cite{Martin:2014sxa}; (iv) a 60 GeV
non-self-conjugate particle annihilating with cross-section
$\avg{\sigma v} = 6.1\times 10^{-26}\mathrm{cm^3/s}$ to 40 GeV scalars
which subsequently decay to pairs of gluons \cite{Martin:2014sxa}.
All photon spectra have been computed using Pythia 8
\cite{Sjostrand:2007gs}.  The spike signals vary by $\sim 50\%$ for
particle models that make nearly identical predictions for the halo
signal.

In summary, if the GC excess is indeed due to DM, an adiabatic spike
at the location of \sgra can only be reconciled with observation if
{\em multiple} input parameters differ significantly from their
central values: for instance, if the velocity dispersion in the Milky
Way is 1$\sigma$ or more higher than the central value $v_0 = 105$
km$/$sec {\em and} the inner NFW halo index is shallower than
indicated by fits, $\gamma_c \lesssim 1.1$ {\em and} the annihilation
density $\rho_{\mathrm{ann}}$ is reduced by a factor of 0.75.  Our PSF
smoothing of $1^\circ$ is a conservative choice. Smaller angular
values, as appropriate for the $\gtrsim 1$ GeV photons in the excess,
would give higher spike/halo ratios and exacerbate the tension between
predictions for the canonical adiabatic spike and observations of the
point source.

As the canonical adiabatic spike is thus disfavored by observation,
other possible scenarios yielding different DM density spikes are
interesting.  The lower two curves in Fig.~\ref{fig:spectrum} show the
flux predictions for shallower spikes, as expected in the presence of
(e.g.) stellar heating. For $\gamma_{sp} = 1.8$, the annihilation
signal is dominated by the spike, while for $\gamma_{sp}=1.5$, the
annihilation signal is mostly due to the halo; contributions from
shallower spikes will be largely overshadowed by the halo emission.
Both scenarios are in accord with the data, and lie intriguingly close
to the observed emission at high energies (1-10 GeV).  At low
energies, where the observed spectrum exceeds the DM prediction, other
astrophysical sources of gamma rays would be expected to contribute to
the observed emission.

The 2nd {\em Fermi} Catalog does not make a firm identification of
2FGL J1745.6-2858, proposing that it may be one of only three pulsar
wind nebulae seen without identified pulsars. Given the spectrum of
the GC excess, a DM origin from a shallow spike $\gamma_{sp}\lesssim
1.8$ is an interesting alternate explanation.  On the other hand, if
future work (e.g., pulsar discovery) does associate 2FGL J1745.6-2858
with an astrophysical source, such as the Sgr A East supernova
remnant, then the case for a DM spike would weaken even more.

\vspace{0.2cm}

\noindent {\it{\bf Is there a Canonical Spike?}}
{\it Fermi} observations do not agree with our predicted signal from a
canonical SMBH spike for DM interpretations of the GC excess.  The
predicted signal exceeds observations by over an order of magnitude.
Plausible alternatives for the spike include: (i) replacing the
assumption of the adiabatic growth of the SMBH from a smaller seed by
more violent processes, such as mergers or cannibalism, or (ii) the
gravitational heating of DM by baryons.  These processes flatten the
spike and thereby reduce emission.  Abandoning DM isotropy in the GC,
which does not arise in simulated or most analytical
halos~\cite{Vas07}, only steepens the profile near $r_{\rm in}$,
worsening the discrepancy.  Particle physics alternatives are perhaps
more extreme, and include: (i) reducing $\rho_{\rm ann}$ while holding
$\avg{\sigma v}/m^2$ fixed, and thereby reducing the spike signal
relative to the halo, although bringing the canonical spike down to
levels compatible with observations by altering $\rho_{\rm ann}$ alone
requires reducing $\rho_{\rm ann}$ by two orders of magnitude; (ii)
arranging for a cancellation between $s$-wave and higher partial wave
processes that becomes relevant for $v\sim 10 v_0$, which does not
seem especially well-motivated in light of the near-thermal values of
the $s$-wave cross-section required to fit the GC excess; or (iii) DM
particle self-interactions. The latter can not only core the NFW halo
but also soften the SMBH spike~\cite{ShaP14} and thus will
significantly modify the spike and halo emission~\cite{nextup}.

Improved angular resolution, as from \cite{gammalight}, would help
clarify the magnitude and spectrum of the point source and
distinguish it from the spatially extended excess.  Other
higher-resolution instruments that can probe the GC environment (e.g., the
{\em Cerenkov Telescope Array} \cite{Consortium:2010bc}, {\it
  GAMMA-400} \cite{gamma400}, the {\it Event Horizon Telescope}
\cite{Lacroix11}, and even {\it Planck} \cite{Lacroix:2013qka}) may further
refine our understanding.

A point-like SMBH spike is an inescapable component of any DM signal
from the GC, and should be incorporated in analyses.
The relative strengths of the spike and halo emission convey
unique information about the Galactic formation history
as well as the particle properties of DM.

\vspace{0.1cm}

\noindent {\it Acknowledgments}.  We are happy to acknowledge useful
conversations with T.~Linden and T.~Slatyer.
This paper was supported in part by NSF Grants PHY-0963136 and
PHY-1300903 as well as NASA Grants NNX13AH44G and NNX10AC86G at the
University of Illinois at Urbana-Champaign.

\end{document}